\documentclass[pre,showkeys,preprintnumbers,amsmath,amssymb,superscriptaddress,twocolumn]{revtex4-2}
\pdfoutput=1 
\usepackage{graphicx}
\usepackage{amsfonts}
\usepackage{url}
\usepackage{epstopdf}
\usepackage{color}
\usepackage{bm}
\usepackage[colorlinks=true,linkcolor=blue,citecolor=red]{hyperref}%

\def\J{{\bf{J}}}
\def\n{\bm{n}}
\def\L{\mathcal L}
\def\N{\mathcal N}

\def\pa{\partial\Omega}
\def\E{{\mathbb E}}
\def\P{{\mathbb P}}
\def\R{{\mathbb R}}

\def\x{\bm{x}}
\def\X{\bm{X}}

\begin{document}

\title{Reaction-Diffusion Processes with Surface Autocatalysis}

\author{Denis~S.~Grebenkov}
 \email{denis.grebenkov@polytechnique.edu}
\affiliation{
Laboratoire de Physique de la Mati\`{e}re Condens\'{e}e, \\ 
CNRS -- Ecole Polytechnique, Institut Polytechnique de Paris, 91120 Palaiseau, France}

\date{\today}

\begin{abstract}
Autocatalytic processes underlie diverse systems in which replication
is triggered at interfaces, including heterogeneous catalysis on solid
substrates, enzyme activity at membranes, viral infections, biofilm
growth, and spatially structured ecosystems.  In a typical scenario,
particles diffuse through a bulk medium and interact with surface
regions, where they may either disappear or reproduce through
branching, cloning, or splitting.  The interplay between loss and
replication at surfaces gives rise to rich population dynamics.  Here
we develop a general theoretical framework for diffusion-mediated
autocatalytic processes at surfaces.  We derive a nonlinear
renewal-type integral equation for the generating function of the
population size, which provides access to its full probability
distribution and integer moments.  We further establish an equivalent
description in terms of a Fokker-Planck equation with nonlinear Robin
boundary conditions that encode surface autocatalytic reactions.  Our
results identify universal asymptotic regimes and provide a unified
framework to predict when surface activity promotes extinction or
explosive growth of the population.  The developed quantitative
framework opens new avenues for analyzing catalytic efficiency,
metabolic regulation, and population persistence in spatially
heterogeneous environments.
\end{abstract}




\keywords{nonlinear dynamics, branching processes, diffusion-mediated phenomena, 
population growth, heterogeneous catalysis, autocatalytic reaction}

\maketitle

\section{Introduction}  

Diffusion-mediated processes play a key role in many physical,
chemical, and biological phenomena, including heterogeneous catalysis
in porous media \cite{Ben-Avraham,Lindenberg}, spin relaxation on
magnetic impurities \cite{Brownstein79,Grebenkov07}, recombination
processes \cite{Sano79,Agmon88}, fluorescence quenching
\cite{Wilemski73}, macromolecular interactions \cite{Berg85},
reactions in micellar and vesicular systems \cite{Sano81},
intracellular transport \cite{Bressloff13,Holcman13,Benichou14},
respiration \cite{Weibel,Sapoval02,Grebenkov05,Serov16}, animal
foraging and search strategies \cite{Oshanin09,Edwards07,Benichou11},
the spread of diseases and epidemics \cite{Goltsev12}.  In a typical
setting, a particle $A$ (e.g., a molecule, an ion, a spin, a quantum
dot, a protein, a virus, a bacterium, an animal) moves through the
surrounding medium, until it encounters a reactive region $\Gamma_0$,
where it may relax its excited state (such as spin magnetization or
fluorophore excitation), be trapped, chemically transformed,
disassembled, disintegrated, passivated, transported outside the
confinement through a channel, eaten or killed \cite{Grebenkov23g}.
Whatever the microscopic origin of the reaction event, the particle is
effectively removed from the system via a reaction pathway $A +
\Gamma_0 \to \Gamma_0$, characterized by a surface reactivity
$\kappa_0$ \cite{Collins49,Grebenkov20,Piazza22}.  Numerous studies
have investigated how the reactivity $\kappa_0$ controls the overall
production rate of a chemical reactor, the functioning of respiratory
organs, the distribution of the associated first-reaction times in
living cells, and efficiency of diffusion-mediated search processes
(see \cite{Redner,Krapivsky,Schuss,Metzler,Grebenkov} and references
therein).

The opposite behavior arises in branching processes, including
autocatalytic (self-amplifying) branching reactions in the bulk.
Under favorable conditions, each particle may split into two or more
offspring, implying a rapid growth of the population over time.
Classical examples range from neutron production in a nuclear reactor
to bacterial colony growth, viral infections, population models in
ecology, and genealogical trees in social systems
\cite{Harris,Athreya,Williams,Kimmel}.  Various types of autocatalytic
reactions play an important role in chemistry and biology
\cite{Plasson11,Bissette13,Blokhuis20,Hanopolskyi21,Laurant08}.  For
instance, autocatalysis as a positive-feedback mechanism appears as an
essential element for controlling complex networks of chemical
reactions \cite{Kriukov24}, whereas the whole concept of
self-replication is crucial for understanding the origins of life
\cite{Oparin,Hordijk13,Preiner19}.  In particular, ribosome biogenesis
provides a prominent example of effective autocatalysis in biological
systems, where the molecular machinery required for protein synthesis
contributes to its own production \cite{Reuveni17}.  The intrinsic
nonlinear character of branching processes is responsible for many
distinctive features in chemical and living systems such as formation
of patterns and traveling waves \cite{Murrey,Miguez07}.

\begin{figure}
\begin{center}
\includegraphics[width=0.46\columnwidth]{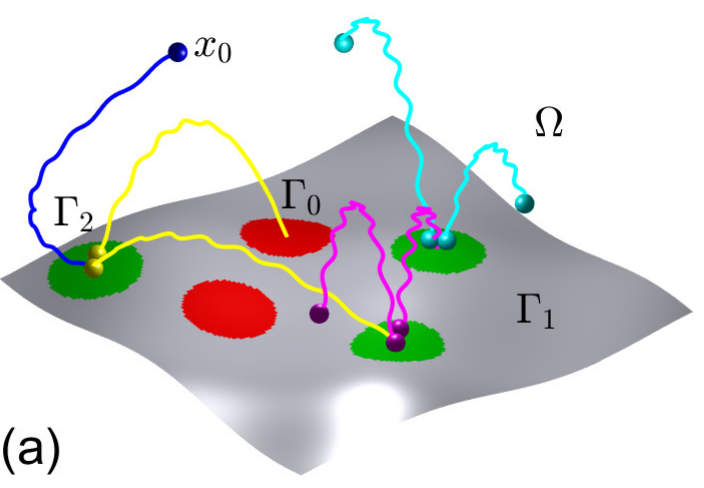} 
\includegraphics[width=0.52\columnwidth]{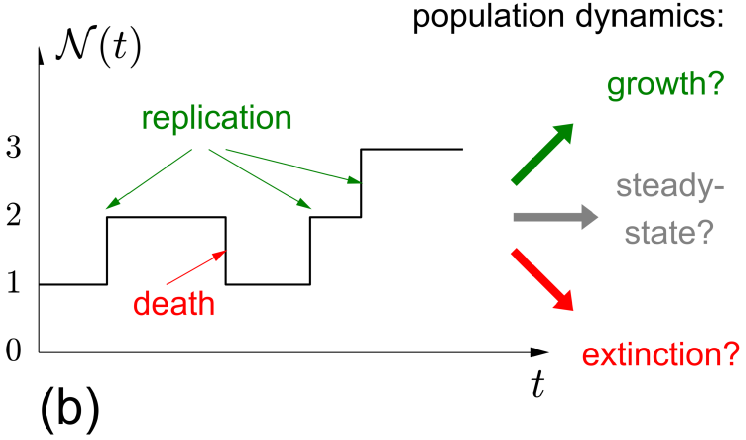} 
\end{center}
\caption{
A schematic view of autocatalytic reactions on a surface partitioned
into three subsets (here, $M = 2$): an absorbing region $\Gamma_0$ (in
red) that destroys particles via $A + \Gamma_0 \to \Gamma_0$; an inert
region $\Gamma_1$ (in gray) that reflects particles back into the
bulk; and a catalytic region $\Gamma_2$ (in green) that replicates
particles via a binary splitting $A + \Gamma_2 \to 2A + \Gamma_2$.
Each of three subsets can be composed of multiple pieces.  In the
shown random realization, a single particle started from $\x_0$
diffuses towards the surface and then branches into two copies on
$\Gamma_2$; one of them reaches an absorbing region $\Gamma_0$ and
disappears, whereas the other produces two more offsprings on another
piece of the catalytic region $\Gamma_2$, and so on.  Competition
between absorbing and branching events determines the stochastic
dynamics of the population size $\N(t)$. }
\label{fig:scheme}
\end{figure}

In this paper, we propose a conceptual extension of the classical
framework of diffusion-controlled reactions
\cite{North66,Calef83,Galanti16,Grebenkov23g} by incorporating
self-amplifying branching events {\it at surfaces.}  In contrast to
previous studies on autocatalytic reactions in the bulk or even under
well-stirred conditions (i.e., without any space dependence), we
consider diffusion-mediated autocatalytic reactions, which are
triggered upon interaction with specific regions on the surface of the
confinement.  More specifically, the boundary of a confining domain
$\Omega$ is supposed to be partitioned in a finite number of regions
$\Gamma_0, \Gamma_1, \cdots, \Gamma_M$, as schematically illustrated
in Fig. \ref{fig:scheme}.  When a particle hits a partially absorbing
region $\Gamma_0$, it may disappear with reactivity $\kappa_0$, as
described above.  Similarly, upon hitting any region $\Gamma_m$ with
$m \geq 2$, the particle may undergo a branching event with reactivity
$\kappa_m$ into $m$ independent identical copies of itself that start
diffusing from the position of the branching event (in this scheme,
the region $\Gamma_1$ just reflects the particle back to $\Omega$).
The conceptual distinction of this model is a unified description of
reaction events that encompasses autocatalytic reactions $A +
\Gamma_m \to mA + \Gamma_m$ of arbitrary order $m$.  If the absorbing
region $\Gamma_0$ is located on the outer confining boundary (e.g., a
plasma membrane of a living cell), it may represent open holes or
channels allowing particles to leave the domain.  In this scenario,
one can describe autocatalytic reactions inside a confinement with
possible leakage.  While the effect of catalytic sites on the
population dynamics was earlier studied for lattice random walks
\cite{Redner84,Albeverio98,Bogachev98,Kesten03,Vatutin04,Vatutin07,Hu12,Topchi13,Bulinskaya18,Bauer21},
a rigorous but physically intuitive introduction of surface
autocatalytic reactions for continuous diffusion processes is one of
the main challenges and central contributions of the present work.

We aim to characterize the population dynamics initiated by a single
particle released from a point $\x_0$ at time $0$.  Since the number
of degrees of freedom of this system evolves randomly, its full
description is challenging and requires elaborate mathematical tools
such as measure-valued stochastic processes (or superprocesses)
\cite{LeGall,Dawson99,Dynkin,Morters05,Englander07}.
We avoid these complications by restricting our attention to the
population size $\N(t)$ -- the number of particles at time $t$.  This
discrete-valued stochastic process is fully described by its
generating function $G_s(t|\x_0) = \E_{\x_0}\{ s^{\N(t)}\}$, with a
parameter $s \in [0,1]$.  Thus, the generating function determines the
probability of having $k$ particles at time $t$, $Q_k(t|\x_0) =
\P_{\x_0}\{ \N(t) = k\} = (\partial_s^k G_s(t|\x_0))_{s=0}/k!$, as
well as the $k$-th order moment: $N_k(t|\x_0) = \E_{\x_0}\{
[\N(t)]^k\} = ((s \partial_s)^k G_s(t|\x_0))_{s=1}$.  In other words,
the knowledge of the generating function gives access to the main
characteristics of the population size.  In particular, the mean
population size, $N_1(t|\x_0)$, generalizes the notion of the survival
probability \cite{Bray13,Levernier19}.

\section{Nonlinear stochastic dynamics}  

The generating function can be obtained from the following renewal
argument based on the first reaction event.  Up to time $t$, the
initial single particle either reacts on one of regions $\Gamma_0,
\cdots,\Gamma_M$, or does not react.  Let $\tau_m$ be the
first-reaction time (FRT) on $\Gamma_m$, provided that the first
reaction event has occurred on $\Gamma_m$.  Using a shorthand notation
$T = \min\{t,\tau_0,\ldots,\tau_M\}$, we represent the generating
function as
\begin{align} \label{eq:Gs_prob}
G_s(t|\x_0) & = \E_{\x_0}\{ s\, 1_{t = T} \} \\  \nonumber
& + \sum\limits_{m=0}^M
\E_{\x_0}\{ s^{\N_1(t-\tau_m) + \cdots + \N_m(t-\tau_m)} 1_{\tau_m = T}\}.
\end{align}
In the first term, the condition $t = T$ means that no reaction event
occurred up to $t$ (i.e., all first-reaction times $\tau_m$ exceed
$t$), and the population size $\N(t)$ is still $1$.  In turn, if
$\tau_m$ is the smallest FRT (and it is below $t$), the reaction event
on $\Gamma_m$ produces $m$ independent particles.  In the remaining
$t-\tau_m$ time, each of these particles can generate its descendants,
with $\N_k(t-\tau_m)$ denoting the random number of particles in the
$k$-th subpopulation ($k = 1,\ldots,m$).  The term with $m = 0$
corresponds to the absorption event, for which there is no descendant,
yielding $s^0 = 1$.

To derive a closed-form equation, we need the joint probability
density $j_m(\x,t|\x_0)$ of the FRT $\tau_m$ and the location
$\X_{\tau_m}$ of the associated reaction event.  This density is given
by the restriction of the probability flux density to $\Gamma_m$,
which can be expressed in terms of the propagator $p(\x,t|\x_0)$ that
describes the probability density of finding the initial particle near
a point $\x$ at time $t$, given that it has not reacted on any of the
regions $\Gamma_0, \ldots, \Gamma_M$.  For a general diffusion
process, the propagator satisfies the Fokker-Planck equation with
Robin-type boundary condition $j_m(\x,t|\x_0) = \kappa_m
p(\x,t|\x_0)|_{\Gamma_m}$ on each $\Gamma_m$ (see Appendix
\ref{sec:model} for more details).  We can therefore rewrite
Eq. (\ref{eq:Gs_prob}) more explicitly as
\begin{align}  \label{eq:Gs_int}
G_s(t|\x_0) & = s + \sum\limits_{m=0}^M \int\limits_{\Gamma_m} d\x  \\  \nonumber
& \int\limits_0^t dt' \kappa_m \, p(\x,t'|\x_0) \bigl([G_s(t-t'|\x)]^m - s\bigr).
\end{align}
This closed nonlinear integral equation, which determines the
generating function $G_s(t|\x_0)$ for an arbitrary diffusion process
with surface autocatalysis, is the first main result.  In this way,
the statistical properties of the randomly evolving system of multiple
particles are described in terms of the single-particle propagator
$p(\x,t|\x_0)$.

An equivalent PDE formulation provides additional physical insight.
For clarity of presentation, we assume below that the particles
undergo ordinary diffusion with a constant diffusivity $D$.  The
extension to general diffusion processes is given in Appendix
\ref{sec:PDE}.  In this setting, the propagator satisfies the
diffusion equation $\partial_t p = D \Delta p$, subject to the initial
condition $p(\x,0|\x_0) = \delta(\x-\x_0)$ prescribing the starting
point $\x_0$ at time $0$, and the Robin boundary condition $-D
\partial_n p = \kappa_m p$ on each $\Gamma_m$ to account for possible
reactions, where $\partial_n$ is the normal derivative oriented
outward from the domain $\Omega$.  The key feature of
Eq. (\ref{eq:Gs_int}) is that the generating function $G_s(t|\x_0)$ in
the bulk is determined as the convolution of the propagator with a
nonlinear boundary term.  The same convolution structure arises in the
representation formula for diffusion equations with inhomogeneous
boundary conditions.  Comparing these two representations (see details
in Appendix \ref{sec:PDE}), we obtain that the generating function
$G_s(t|\x_0)$ satisfies the (backward) diffusion equation,
\begin{subequations}
\begin{equation}  \label{eq:Gs_diff}
\partial_t G_s(t|\x_0) = D\Delta G_s(t|\x_0) ,
\end{equation}
subject to the initial condition $G_s(0|\x_0)= s$ and the boundary
condition on each region $\Gamma_m$:
\begin{equation}  \label{eq:Gs_BC}
D \partial_n G_s(t|\x_0) = \kappa_m \bigl( [G_s(t|\x_0)]^m - G_s(t|\x_0)\bigr).
\end{equation}
\end{subequations}
Equation (\ref{eq:Gs_BC}) is a nonlinear generalization of the
classical Collins-Kimball Robin boundary condition and constitutes the
central theoretical result of this work.  Remarkably, all nonlinearity
is confined to the boundary condition, whereas the diffusion equation
in the bulk remains linear.

The initial condition reflects the presence of a single particle at
time $0$, $\N(0) = 1$, and follows immediately from
Eq. (\ref{eq:Gs_int}).  As the population size does not change in the
bulk, one naturally recovers the standard diffusion equation
(\ref{eq:Gs_diff}).  In turn, the {\it nonlinear} Robin boundary
condition (\ref{eq:Gs_BC}) encodes the effect of branching events
occurring at the boundary and is the defining distinction from
conventional diffusion-controlled reactions.  Physically,
Eq. (\ref{eq:Gs_BC}) reflects how the generating function changes from
$G_s(t|\x_0)$ to $[G_s(t|\x_0)]^m$ upon the autocatalytic reaction $A
+ \Gamma_m \to m A + \Gamma_m$.  As the reaction occurs at the
boundary, this change is not associated with an increment of physical
time but with an increment of the boundary local time, which naturally
gives rise to the normal derivative (for mathematical details, see,
e.g., \cite{Delmas05,Barbu16}).  The right-hand side of
Eq. (\ref{eq:Gs_BC}) vanishes identically for $m = 1$.  Accordingly,
the reaction $A + \Gamma_1 \to A + \Gamma_1$ leaves the population
unchanged so that $\Gamma_1$ represents an inert boundary region, and
the value of $\kappa_1$ is irrelevant.

The above framework provides a simple qualitative picture of the
population dynamics.  If there is no absorption (i.e., $\kappa_0 =
0$), branching events occur in parallel for independent particles and
lead to an exponential growth of the population size.  In general,
however, branching and absorption events compete with each other, and
the dynamics of the system strongly depends on the geometric structure
of the confinement, on the partition of its boundary into the regions
$\Gamma_0,\ldots,\Gamma_M$, and on their reactivities
$\kappa_0,\ldots,\kappa_M$.  While we are not aware of any explicit
solution of the integral equation (\ref{eq:Gs_int}) even for simple
geometries (e.g., an interval), its long-time behavior can be
understood in a general situation.

We begin by analyzing the mean population size $N_1(t|\x_0)$, which is
obtained by evaluating the first derivative of $G_s(t|\x_0)$ at $s =
1$ and thus obeys
\begin{subequations}  \label{eq:N1}
\begin{align} \label{eq:N1_diff}
\partial_t N_1(t|\x_0) &= D\Delta N_1(t|\x_0) \quad (\x_0\in \Omega), \\  \label{eq:N1_BC}
D \partial_n N_1(t|\x_0) &= (m-1) \kappa_m N_1(t|\x_0) \quad (\x_0\in \Gamma_m), \\  \label{eq:N1_ini}
N_1(0|\x_0) &= 1.
\end{align}
\end{subequations}
Remarkably, despite the nonlinear character of branching events, the
problem reduces to a {\it linear} boundary-value problem for
$N_1(t|\x_0)$ that can hence be analyzed by conventional spectral tools
\cite{DelGrosso76,Grebenkov26a}.  On the partially absorbing region 
$\Gamma_0$, the Robin boundary condition states that the diffusive
flux density of particles from the bulk, $(-D\partial_n
N_1)|_{\Gamma_0}$, is equal to $\kappa_0 N_1$ -- the flux density of
absorbed particles on $\Gamma_0$.  In contrast, the autocatalytic
reaction $A + \Gamma_m \to m A + \Gamma_m$ on $\Gamma_m$ replaces one
particle by $m$ identical copies, so that the diffusive flux density
$(-D\partial_n N_1)|_{\Gamma_m}$ is equal to $-(m-1) \kappa_m N_1$,
and is consequently directed from the boundary into the bulk.  The
effect of the autocatalytic reaction on the mean population size is
therefore equivalent to a {\it negative} surface reactivity
$-(m-1)\kappa_m$ on $\Gamma_m$.  The time evolution of the mean
population size is consequently governed by the competition between
regions with positive and negative reactivities.
The spectral decomposition of Eqs. (\ref{eq:N1}) yields the long-time
behavior $N_1(t|\x_0) \propto e^{-\lambda_0 t}$, which is controlled
by the principal (smallest) eigenvalue $\lambda_0$ of the governing
diffusion operator $-D\Delta$ or, more generally, of the Fokker-Planck
operator (see Appendix \ref{sec:model}).  The crucial difference from
conventional diffusion-controlled reactions is that $\lambda_0$ can
now be positive, zero, or negative, thus defining subcritical,
critical, and supercritical regimes, respectively.  The mean
population size can vanish exponentially ($\lambda_0 > 0$), reach a
steady-state limit ($\lambda_0 = 0$), or grow exponentially
($\lambda_0 < 0$).  In the following, we uncover the long-time
behavior of the population size beyond its mean value.

Higher-order moments $N_k(t|\x_0)$, as well as the probabilities
$Q_k(t|\x_0)$, are obtained from the generating function $G_s(t|\x_0)$
by evaluating $k$-fold derivatives with respect to $s$.  In this way,
one can either deduce integral equations for $Q_k(t|\x_0)$ and
$N_k(t|\x_0)$, in analogy to Eq. (\ref{eq:Gs_int}), or a PDE
formulation.  In the latter case, since the derivative with respect to
$s$ does not alter Eq. (\ref{eq:Gs_diff}), both $Q_k(t|\x_0)$ and
$N_k(t|\x_0)$ obey the (backward) diffusion equation, with the initial
conditions $Q_k(0|\x_0) = \delta_{k,1}$ and $N_k(0|\x_0) = 1$,
respectively.  Likewise, using standard combinatorial relations to
evaluate the $k$-fold derivatives of Eq. (\ref{eq:Gs_BC}), we obtain
the boundary conditions on each $\Gamma_m$ (see Appendices
\ref{sec:Qk} and \ref{sec:Nk}):
\begin{align}  \label{eq:Qn_BC}
& D \partial_n Q_k(t|\x_0) + \kappa_m Q_k(t|\x_0) \\  \nonumber
& \qquad = \kappa_m \sum\limits_{i_1,\ldots, i_m \geq 0 \atop i_1+\ldots+i_m=k}   
Q_{i_1}(t|\x_0) \cdots Q_{i_m}(t|\x_0)
\end{align}
and
\begin{align}  \label{eq:Nk_BC}
& D \partial_n N_k(t|\x_0) + \kappa_m N_k(t|\x_0) \\  \nonumber
& \qquad = \kappa_m \sum\limits_{i_1,\ldots, i_m \geq 0 \atop i_1+\ldots+i_m=k}   
\frac{k!}{i_1! \cdots i_m!} N_{i_1}(t|\x_0) \cdots N_{i_m}(t|\x_0),
\end{align}
with $N_0(t|\x_0) \equiv 1$.  This is the third main result.  

In particular, the probability $Q_0(t|\x_0)$ of having no particle at
time $t$ satisfies the nonlinear boundary condition
\begin{align}  \label{eq:Q0_BC}
& D \partial_n Q_0(t|\x_0) = \kappa_m \bigl[Q_0^m(t|\x_0) - Q_0(t|\x_0)\bigr] \quad (\x_0\in\Gamma_m).
\end{align}
In analogy to the survival probability for conventional
diffusion-controlled reactions, the time derivative of $Q_0(t|\x_0)$
determines the probability density of the population extinction time
-- the key quantity in many applications.  Once $Q_0(t|\x_0)$ is
computed, the other probabilities $Q_k(t|\x_0)$ (with $k \geq 1$) can
be found iteratively by solving the (backward) diffusion equation with
the boundary condition (\ref{eq:Qn_BC}), which is {\it linear} in
$Q_k(t|\x_0)$.  Likewise, the boundary condition (\ref{eq:Nk_BC}) is
linear in $N_k(t|\x_0)$, i.e., each moment satisfies a linear
boundary-value problem once all lower-order moments are known.
However, the essential difference to the linear problem (\ref{eq:N1})
is that the product in the right-hand side of Eqs. (\ref{eq:Qn_BC},
\ref{eq:Nk_BC}) renders standard techniques such as Laplace transform
inefficient.

\section{Three asymptotic regimes}  

A quantitative description of the population dynamics requires either
an elaborate asymptotic analysis or a numerical solution of the above
integral or differential equations that will be reported in a separate
publication.  Conversely, the qualitative picture can be drawn from
the general structure of these equations.  As earlier, we distinguish
three regimes.

(i) In the subcritical regime ($\lambda_0 > 0$), the absorption events
on $\Gamma_0$ progressively eliminate all the particles that were
produced via autocatalytic reactions on $\Gamma_2, \ldots,
\Gamma_M$.  As a result, the mean population size, as well as
higher-order moments $N_k(t|\x_0)$ and the probabilities $Q_k(t|\x_0)$
with $k > 0$ vanish exponentially at long times.  Moreover, the decay
rates of all these quantities are expected to be equal to $\lambda_0$,
as for the mean population size.  In other words, the spectral
solution of the linear problem (\ref{eq:N1}) universally controls the
long-time behavior of the whole population dynamics.  In this regime,
the extinction of the population resembles an exponential decrease of
the concentration of particles in conventional diffusion-controlled
reactions in bounded domains.

(ii) The behavior is drastically different in the supercritical regime
($\lambda_0 < 0$) when the absorption events are not frequent enough
to compensate for the proliferative effect of branching events, so
that more and more particles are produced, yielding an exponential
growth $N_1(t|\x_0) \propto e^{|\lambda_0|t}$, with the rate
$|\lambda_0|$.  As the right-hand side of Eq. (\ref{eq:Nk_BC})
involves products of $N_{i_1}, \ldots, N_{i_m}$ such that $i_1 +
\ldots + i_m = k$, the $k$-th moment $N_k(t|\x_0)$ grows as
$e^{k|\lambda_0|t}$, with its own rate $k|\lambda_0|$.  This
asymptotic behavior suggests that the rescaled population size,
$\N(t)/N_1(t|\x_0)$, may approach a steady-state distribution at long
times.  In other words, the population dynamics is essentially
controlled by the mean population size.  Additional insight can be
gained by inspecting the behavior of the probabilities $Q_k(t|\x_0)$.
For any fixed $k \geq 2$, the probability $Q_k(t|\x_0)$ increases at
short times, reaches a maximum, and then asymptotically vanishes at
long times.  In fact, as the population grows exponentially, it is
less and less likely to observe a fixed number of particles as
$t\to\infty$.  However, the decay of $Q_k(t|\x_0)$ is not necessarily
exponential that makes the supercritical regime more intricate.

\begin{figure}
\begin{center}
\includegraphics[width=0.49\columnwidth]{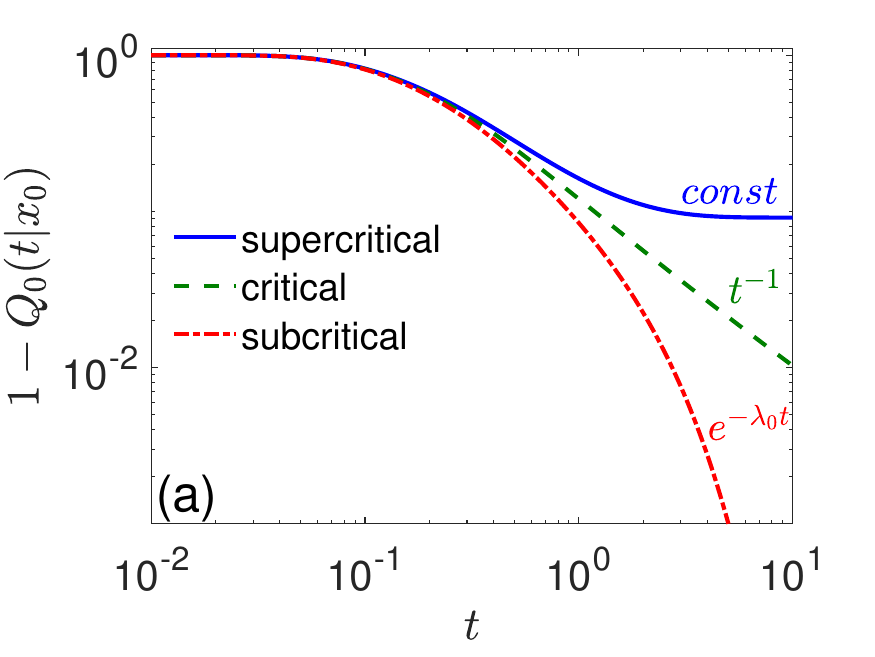} 
\includegraphics[width=0.49\columnwidth]{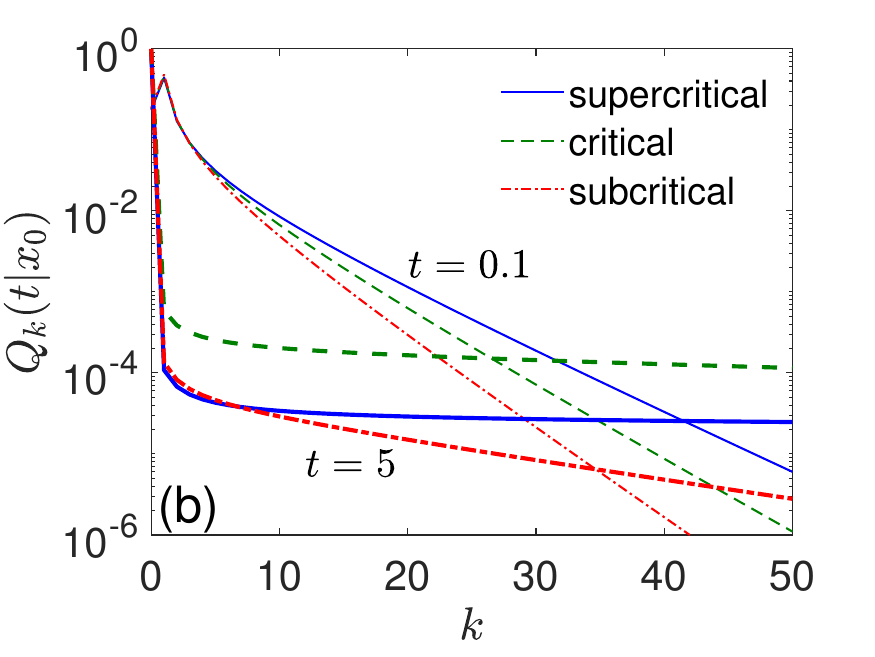} 
\includegraphics[width=0.49\columnwidth]{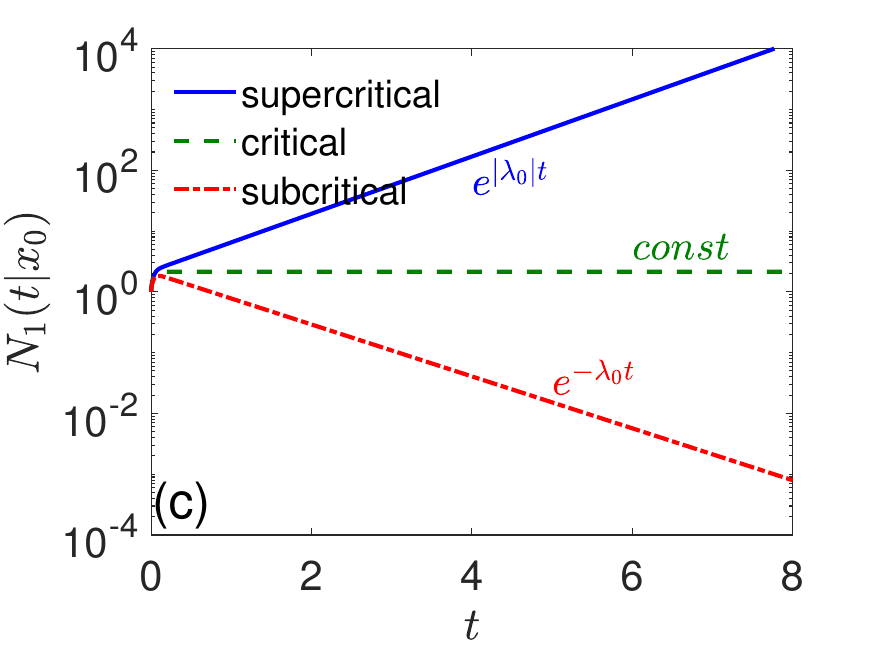} 
\includegraphics[width=0.49\columnwidth]{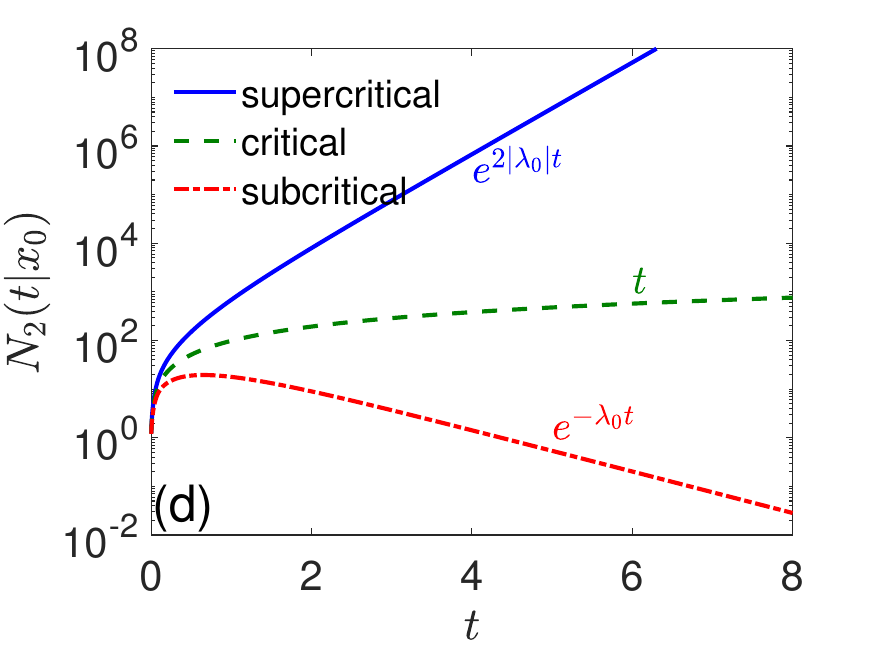} 
\end{center}
\caption{
Time evolution of the population size $\N(t)$ for a diffusion-reaction
system with diffusivity $D = 1$ in a hollow cylinder of radii $R =
0.2$ and $L = 1$, with the outer perfectly absorbing surface
($\kappa_0 = \infty$) and the inner autocatalytic surface with
reactivity $\kappa_2$.  The starting point $\x_0$ is on the inner
surface.  Three regimes are shown for $\kappa_2 = 0.9\, \kappa_2^c$
(subcritical, dash-dotted line), $\kappa_2 = \kappa_2^c$ (critical,
dashed line), and $\kappa_2 = 1.1\, \kappa_2^c$ (supercritical, solid
line), with $\kappa_2^c = D/(R\ln(L/R)) \approx 3.11$
\cite{Grebenkov26a}.  Arbitrary units are used.  {\bf (a)} Probability
$1-Q_0(t|\x_0)$ of having at least one particle at time $t$; {\bf (b)}
Distribution $Q_k(t|\x_0)$ at two times: $t = 0.1$ (thin lines) and $t
= 5$ (thick lines); {\bf (c)} Mean population size $N_1(t|\x_0)$; {\bf
(d)} Second-order moment $N_2(t|\x_0)$. }
\label{fig:Qk}
\end{figure}

(iii) Expectedly, the critical regime ($\lambda_0 = 0$) exhibits the
most peculiar features.  While the mean population size reaches a
steady-state limit, higher-order moments of $\N(t)$ grow with time.
In fact, even though the reactivities $\kappa_m$ are tuned to
compensate the opposite effects of absorption and branching events
{\it on average} (for $N_1(t|\x_0)$), this balance is purely
statistical.  In other words, the population size $\N(t)$ does not
fluctuate around the mean value $N_1(t|\x_0)$, in analogy to, say, an
Ornstein-Uhlenbeck process.  On the contrary, most random realizations
of the population dynamics vanish such that the probability
$1-Q_0(t|\x_0)$ of having at least one particle at $t$ goes to $0$ at
long times, and the steady-state limit of $N_1(t|\x_0)$ is ensured by
fewer and fewer realizations that managed to produce more and more
particles.  While the relative contributions of the majority of
extinct populations and few survived populations are matched to
achieve a nontrivial limit $N_1(\infty|\x_0)$, the survived
realizations dominate in higher-order moments and lead to their
divergence as $t\to\infty$.  This qualitative argument can be
supported by the analysis of the above PDE description.

In order to illustrate this qualitative picture, we solved the
integral equations for $Q_k(t|\x_0)$ and $N_k(t|\x_0)$ numerically for
the population dynamics inside a hollow cylinder, in which the inner
surface $\Gamma_2$ is catalytic with reactivity $\kappa_2$ and the
outer surface $\Gamma_0$ is perfectly absorbing with reactivity
$\kappa_0 = \infty$.  The rotational invariance of this domain
facilitates the computation of the single-particle propagator
$p(\x,t|\x_0)$ and eliminates the spatial integral over $d\x$ in
Eq. (\ref{eq:Gs_int}) and related integral equations.  Figure
\ref{fig:Qk} illustrates the main features of the population dynamics.
Panel 'a' shows how the probability $1-Q_0(t|\x_0)$ decreases over
time, with an exponential decay in the subcritical regime, a much
slower, power-law decay $t^{-1}$ in the critical regime, and a nonzero
limit in the supercritical regime.  The time evolution of the whole
distribution $Q_k(t|\x_0)$ is shown on panel 'b'.  Starting from a
distribution peaked around $k = 1$ at short times, one sees how the
probability $Q_0(t|\x_0)$ becomes dominant, while all other
$Q_k(t|\x_0)$ progressively decrease.  The shape of the distribution
at long times strongly depends on the asymptotic regimes, with a rapid
decrease of $Q_k(t|\x_0)$ with $k$ in the subcritical regime, and a
much slower decrease in the critical and supercritical regimes, at
least for the shown range of $k$.  This slow decrease explains the
long-time behavior of the mean population size $N_1(t|\x_0)$ (panel
'c') and of the second-order moment $N_2(t|\x_0)$ (panel 'd').

\begin{figure*}
\begin{center}
\includegraphics[width=0.32\textwidth]{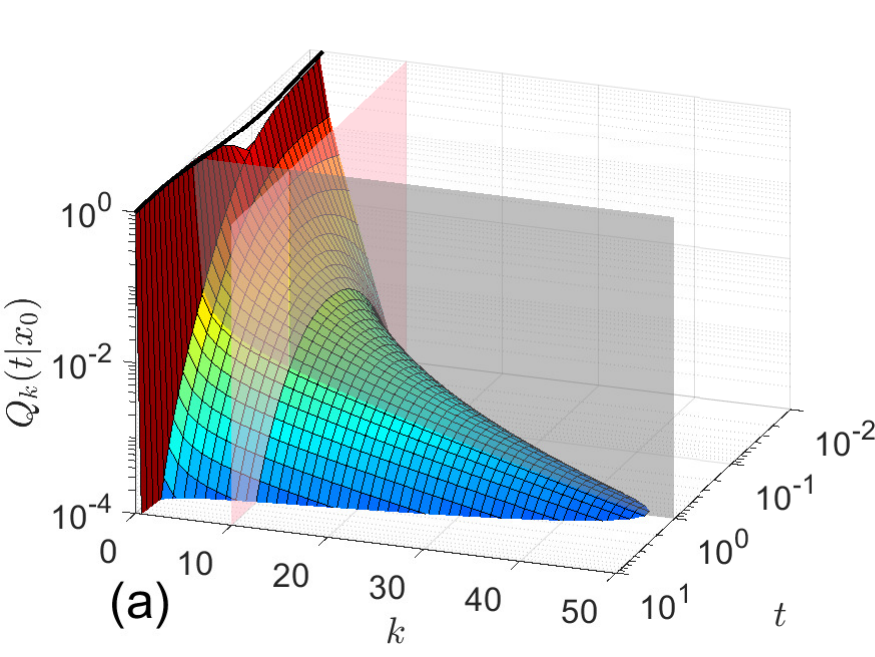} 
\includegraphics[width=0.32\textwidth]{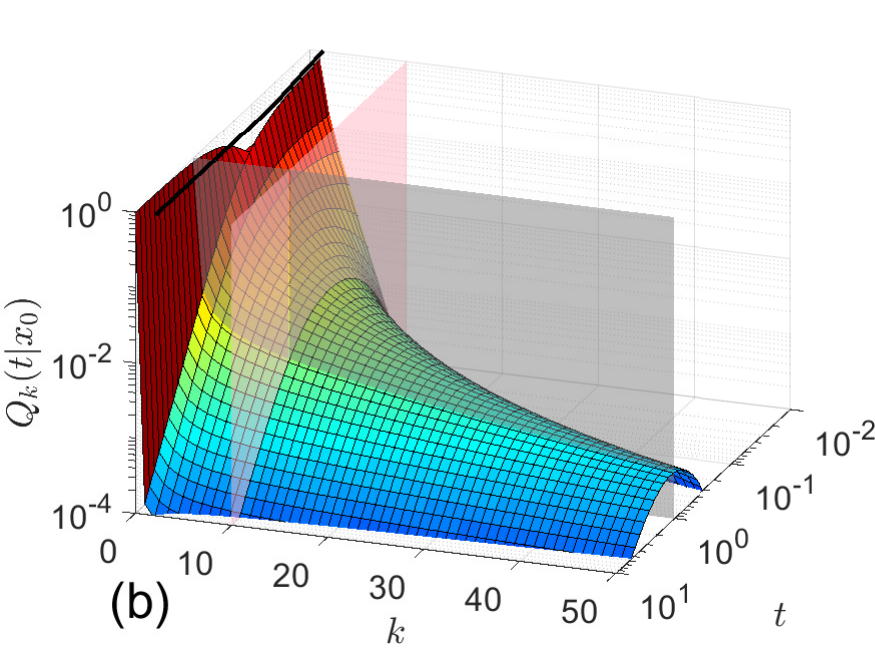} 
\includegraphics[width=0.32\textwidth]{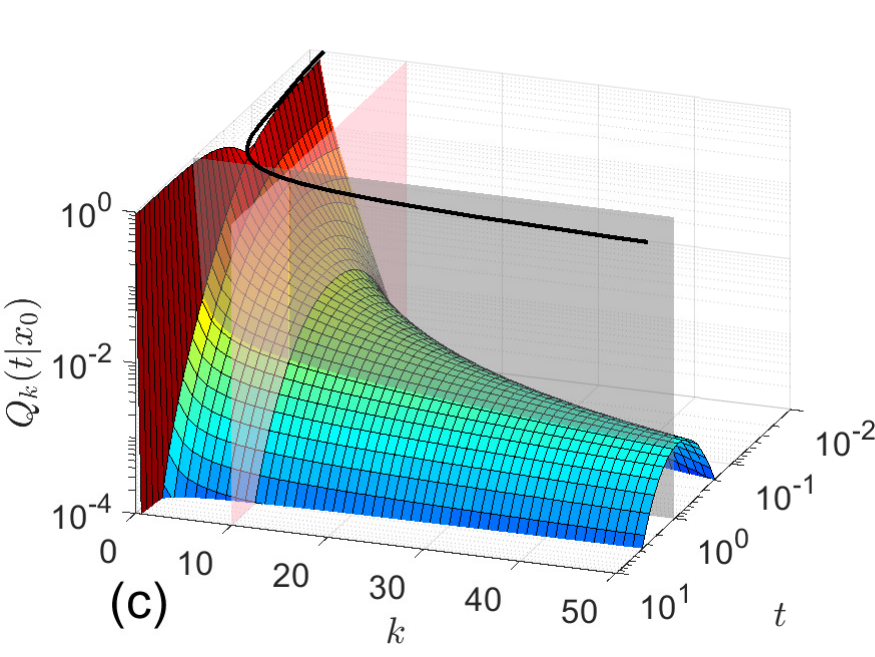} 
\end{center}
\caption{
Time evolution of the population size $\N(t)$ for a diffusion-reaction
system with diffusivity $D = 1$ in a hollow cylinder of radii $R =
0.2$ and $L = 1$, with the outer perfectly absorbing surface
($\kappa_0 = \infty$) and the inner autocatalytic surface with
reactivity $\kappa_2$.  The starting point $\x_0$ is on the inner
surface.  Probabilities $Q_k(t|\x_0)$ of having $k$ particles at time
$t$ for three regimes: {\bf (a)} subcritical ($\kappa_2 = 0.9\,
\kappa_2^c$); {\bf (b)} critical ($\kappa_2 = \kappa_2^c$); and {\bf
(c)} supercritical ($\kappa_2 = 1.1\, \kappa_2^c$), with $\kappa_2^c =
D/(R\ln(L/R)) \approx 3.11$ \cite{Grebenkov26a}.  Arbitrary units are
used.  The black curve indicates the mean population size
$N_1(t|\x_0)$ as a function of time, plotted in the horizontal plane.
Gray and rose cross-sections are added at $t = 1$ and $k = 10$ to
guide eyes for comparison between three regimes. }
\label{fig:Qn_surf2}
\end{figure*}

Figure \ref{fig:Qn_surf2} provides additional insights onto the time
evolution of the distribution $Q_k(t|\x_0)$ in three regimes.  As
three chosen values of $\kappa_2$ differ by only $10\%$, the three
distributions visually look similar, though their long-time asymptotic
properties are drastically different, as seen by the behavior of the
mean population size (black curve).
At short times, the distribution $Q_k(t|\x_0)$ is peaked at $k = 1$
because the population was initiated with a single particle.  As $t$
increases, the probability of having $k \ne 1$ particles increases,
whereas the dominance of $Q_1(t|\x_0)$ declines.  At intermediate
times, the probability $Q_0(t|\x_0)$ of having no particle becomes
dominant, while all other probabilities $Q_k(t|\x_0)$ progressively
decrease.  The major distinctions between three regimes appear at long
times.  In particular, it is highly unlikely to observe a large
population of particles in the subcritical regime (see a rapid decay
of the probability $Q_k(t|\x_0)$ with $k$ at $t = 1$, as illustrated
by gray slice).  In turn, such realizations are more frequent in
critical and supercritical regimes.  In the latter case, the
exponential growth of the mean population size and higher-order
moments is ensured by higher probability of getting very large
populations due to slow decay of $Q_k(t|\x_0)$ with $k$.
These results demonstrate that the transition between subcritical and
supercritical behavior is not merely reflected in the mean population
size but fundamentally reshapes the entire probability distribution.

\section{Conclusion}

In summary, we have developed a theoretical framework that bridges two
previously separate areas: diffusion-controlled reactions at surfaces
and branching processes in the bulk.  While both topics have been
intensively studied for many decades, their common features remain
largely unexplored.  We introduced a general model of autocatalytic
reactions that occur exclusively on surface regions.  Using
probabilistic renewal arguments, we established a nonlinear integral
equation for the generating function that fully characterizes the
statistics and time evolution of the population size.  This equation
links the diffusive motion of an individual particle, represented by
its propagator, to the collective dynamics of the entire
reaction-diffusion system.  An equivalent reformulation in terms of
the backward Fokker-Planck equation with nonlinear Robin boundary
conditions was derived.  Quantifying the competition between the
opposite effects of absorbing and branching events, we identified
three universal asymptotic regimes of autocatalytic dynamics.  An
extension to spatially delimited subpopulations is presented in
Appendix \ref{sec:extension}.

Beyond the specific examples discussed here, the present framework
applies to a broad class of diffusion-mediated autocatalytic systems
at surfaces in physics, chemistry, and biology.  In heterogeneous
catalysis, many industrially relevant catalytic reactions occur at
active sites on porous materials, providing a theoretical basis for
optimizing catalytic efficiency through the geometric design of
surfaces and the spatial organization of reactive regions.  Similar
ideas may also prove useful for controlled drug delivery, where
reactions localized at interfaces regulate release rates.  In
biological contexts, viral infection and intracellular replication can
be viewed as autocatalytic processes shaped by stochastic transport
inside and between cells.  Microbial biofilms present another example
of surface-mediated autocatalysis, where cells transported by a fluid
attach to interfaces and proliferate while competing with detachment
and death.  Our framework provides a common mathematical language for
analyzing, predicting, and ultimately controlling surface-driven
autocatalytic dynamics in complex natural and engineered systems.
More broadly, the nonlinear Robin boundary condition extends the
classical theory of diffusion-controlled reactions to self-amplifying
interfacial processes, providing a general foundation for future
studies of diffusion-mediated autocatalytic dynamics.

\section*{Data availability}

The data that support the findings of this study are available
from the corresponding author upon reasonable request.

\appendix
\section{Theoretical framework for a single particle}
\label{sec:model}

We consider a general diffusion process with a drift vector $\mu(\x)$
and a diffusion matrix $D(\x)$, which are assumed to be
time-independent and sufficiently smooth functions of the spatial
variable $\x$.  The particle diffuses in a bounded domain $\Omega
\subset \R^d$ with a piecewise smooth boundary $\pa$, which is
partitioned into $M+1$ regions: $\pa = \Gamma_0 \cup \ldots \cup
\Gamma_M$.  The propagator for a single particle, $p(\x,t|\x_0)$,
satisfies the forward Fokker-Planck equation for any fixed starting
point $\x_0\in\Omega$ \cite{Risken,Gardiner}:
\begin{subequations}
\begin{align}  \label{eq:propagator_diff}
\partial_t p(\x,t|\x_0) & = \L_{\x} p(\x,t|\x_0)  \quad (\x\in\Omega), \\  \label{eq:propagator_BC}
\n_{\x} \cdot \J(\x,t|\x_0) & = \kappa_m p(\x,t|\x_0) \quad (\x\in \Gamma_m), \\  \label{eq:propagator_ini}
p(\x,0|\x_0) & = \delta(\x - \x_0),
\end{align}
\end{subequations}
where $\n_{\x}$ is the unit normal vector at a boundary point $\x$,
which is oriented outwards the domain $\Omega$, 
\begin{equation}
J_i(\x,t|\x_0) = \mu_i(\x) p(\x,t|\x_0) - \sum\limits_{j=1}^d \partial_{x_j} \bigl(D_{i,j}(\x) p(\x,t|\x_0)\bigr)
\end{equation}
are the components of the probability density flux $\J(\x,t|\x_0)$,
and $\L_{\x}$ is the Fokker-Planck operator acting on a function
$f(\x)$ as
\begin{equation}
\L_{\x} f = - \sum\limits_{i=1}^d \partial_{x_i} \bigl(\mu_i(\x) f\bigr) 
+ \sum\limits_{i,j=1}^d  \partial_{x_i} \partial_{x_j} \bigl(D_{i,j}(\x) f)
\end{equation}
(here we employ the It\^o stochastic convention).  The propagator
$p(\x,t|\x_0)$ describes the likelihood that a particle started from
$\x_0$ at time $0$ is found in a vicinity of $\x$ at time $t$, given
that it has not reacted on the boundary up to time $t$.

The boundary condition (\ref{eq:propagator_BC}) states that the
probability flux from the bulk in the normal direction to the boundary
is equal at any point $\x$ on $\Gamma_m$ to the probability flux of
reacted particles, given by $\kappa_m p(\x,t|\x_0)$, with $\kappa_m
\geq 0$ being the surface reactivity of $\Gamma_m$.  In this
formulation, all regions $\Gamma_m$ with $m\ne 1$ are considered as
partially absorbing because the propagator $p(\x,t|\x_0)$ is used in
the main text to determine the joint probability density
$j_m(\x,t|\x_0)$ and hence to account for the competition between
different boundary regions for the first reaction of a diffusing
particle.  In other words, the propagator describes conventional
diffusion-controlled reactions $A + \Gamma_m \to \Gamma_m$ for all
regions $\Gamma_m$ with $m \ne 1$; conversely, autocatalytic reactions
are implemented via the renewal equation (\ref{eq:Gs_int}), as
explained in the main text.

Rewriting Eq. (\ref{eq:propagator_diff}) as the continuity equation
(with time $t'$ instead of $t$),
\begin{equation*}
\partial_{t'} p(\x,t'|\x_0) = - \nabla \cdot \J(\x,t'|\x_0),
\end{equation*}
integrating it over $\x\in\Omega$, using the divergence theorem and
the boundary condition (\ref{eq:propagator_BC}), one obtains
\begin{align*}
\partial_{t'} \int\limits_{\Omega} d\x \, p(\x,t'|\x_0) & = - \int\limits_{\pa} d\x \, \n_{\x} \cdot\J(\x,t'|\x_0) \\
&  = - \sum\limits_{m=0}^M \kappa_m \int\limits_{\Gamma_m} d\x \, p(\x,t'|\x_0).
\end{align*}
Integrating this identity over $t'$ from $0$ to $t$ yields
\begin{equation}  \label{eq:p_int}
\int\limits_{\Omega} d\x \, p(\x,t|\x_0) = 1 
- \sum\limits_{m=0}^M \kappa_m \int\limits_0^t dt' \int\limits_{\Gamma_m} d\x \, p(\x,t'|\x_0).
\end{equation}

We recall that the propagator $p(\x,t|\x_0)$ also satisfies the
backward Fokker-Planck (or Kolmogorov) equation for any fixed
$\x\in\Omega$ \cite{Risken,Gardiner}
\begin{subequations}
\begin{align}  \label{eq:propagator_BFP}
\partial_t p(\x,t|\x_0) - \L^*_{\x_0} p(\x,t|\x_0) & = 0   \quad (\x_0\in\Omega), \\
\n_{\x_0} \cdot D(\x_0) \nabla_{\x_0} p(\x,t|\x_0) + \kappa_m p(\x,t|\x_0) & = 0 \quad (\x_0\in \Gamma_m), \\  \label{eq:propagator_BFP_ini}
p(\x,0|\x_0) & = \delta(\x - \x_0),
\end{align}
\end{subequations}
where $\L^*_{\x_0}$ is the adjoint Fokker-Planck operator acting on a
function $f(\x_0)$ as
\begin{equation}
\L^*_{\x_0} f = \sum\limits_{i=1}^d \mu_i(\x_0) \partial_{x_{0,i}} f + \sum\limits_{i,j=1}^d D_{i,j}(\x_0) \partial_{x_{0,i}} \partial_{x_{0,j}} f .
\end{equation}
Since the drift and the diffusion matrix do not depend on time, we
employed the forward form in Eq. (\ref{eq:propagator_BFP}) and imposed
the initial condition, instead of using the equivalent backward form
with a terminal condition.

We also recall that the propagator $p(\x,t|\x_0)$ allows one to
represent the solution $U(t'|\x)$ of an inhomogeneous initial-value
problem
\begin{subequations}  \label{eq:U_problem}
\begin{align}  \label{eq:U_diff}
\partial_{t'} U(t'|\x) - \L^*_{\x} U(t'|\x) & = F(t'|\x) \quad (\x\in \Omega), \\
\n_{\x} \cdot D(\x) \nabla_{\x} U(t'|\x) + \kappa_m U(t'|\x) & = f_m(t'|\x) \quad (\x\in \Gamma_m), \\  
U(0|\x) & = U_0(\x),
\end{align}
\end{subequations}
where $F(t'|\x)$, $f_m(t'|\x)$ and $U_0(\x)$ are given functions
(while we wrote $\x$ instead of $\x_0$ and $t'$ instead of $t$, this
is still a backward Fokker-Planck problem).  For this purpose, let us
first rewrite Eq. (\ref{eq:propagator_diff}) as
\begin{equation*}
\partial_{t'} p(\x,t-t'|\x_0) + \L_{\x} p(\x,t-t'|\x_0) = 0  \quad (\x\in\Omega).
\end{equation*}
Multiplying this equation by $U(t'|\x)$, multiplying
Eq. (\ref{eq:U_diff}) by $p(\x,t-t'|\x_0)$, adding these equations,
integrating over $\x\in\Omega$ and using the Green's formula, we have
\begin{widetext}
\begin{align*}
& \int\limits_{\Omega} d\x \, F(t'|\x) p(\x,t-t'|\x_0) 
= \int\limits_{\Omega} d\x \biggl[\bigl(\partial_{t'} U - \L^*_{\x} U \bigr) p(\x,t-t'|\x_0) 
+ \bigl(\partial_{t'} p(\x,t-t'|\x_0) + \L_{\x} p(\x,t-t'|\x_0)\bigr) U(t'|\x) \biggr] \\
& =  \int\limits_{\Omega} d\x \, \partial_{t'} \bigl(U(t'|\x) p(\x,t-t'|\x_0)\bigr) 
+ \int\limits_{\Omega} d\x  \biggl[U(t'|\x) \L_{\x} p(\x,t-t'|\x_0) - p(\x,t-t'|\x_0) \L^*_{\x} U(t'|\x)\biggr] \\
& = \partial_{t'} \int\limits_{\Omega} d\x \bigl(U(t'|\x) p(\x,t-t'|\x_0)\bigr) 
- \sum\limits_{m=0}^M \int\limits_{\Gamma_m} d\x \, 
\biggl[U(t'|\x) \underbrace{\bigl(\n_{\x} \cdot {\bf J}(\x,t-t'|\x_0)\bigr)}_{=\kappa_m p(\x,t-t'|\x_0)}
+ p(\x,t-t'|\x_0) \underbrace{\bigl(\n_{\x} \cdot D(\x) \nabla_{\x} U(t'|\x) \bigr)}_{= f_m(t'|\x) - \kappa_m U(t'|\x)}\biggr] \\
& = \partial_{t'} \int\limits_{\Omega} d\x \bigl(U(t'|\x) p(\x,t-t'|\x_0)\bigr) 
- \sum\limits_{m=0}^M \int\limits_{\Gamma_m} d\x \, p(\x,t-t'|\x_0) \, f_m(t'|\x) .
\end{align*}
\end{widetext}
Integrating this equation over $t'$ from $0$ to $t$, one obtains a
general representation of the solution of Eqs. (\ref{eq:U_problem}) in
terms of the propagator:
\begin{align} \nonumber
U(t|\x_0) & = \int\limits_\Omega d\x \, U_0(\x) \, p(\x,t|\x_0) \\ \nonumber
& + \int\limits_0^t dt' \biggl[\int\limits_{\Omega} d\x \, F(t'|\x) p(\x,t-t'|\x_0) \\   \label{eq:U_general}
& + \sum\limits_{m=0}^M \int\limits_{\Gamma_m} d\x \, f_m(t'|\x) \, p(\x,t-t'|\x_0) \biggr].
\end{align}

\section{Derivation of the PDE}
\label{sec:PDE}

We use the general representation (\ref{eq:U_general}) to show that
the generating function $G_s(t|\x_0)$ satisfies
Eqs. (\ref{eq:U_problem}) with appropriate functions $U_0$, $F$ and
$f_m$.  Since $\N(0) = 1$, the initial condition is $G_s(0|\x_0) = s$.
Setting $U_0(\x) = s$ and $F = 0$ in Eq. (\ref{eq:U_general}) leads to
\begin{equation}  
U(t|\x_0) = s + \sum\limits_{m=0}^M \hspace*{-0.5mm} \int\limits_0^t dt' 
\hspace*{-1mm}  \int\limits_{\Gamma_m} \hspace*{-1mm}  d\x  \bigl(f_m(t'|\x) - s\kappa_m \bigr)  p(\x,t-t'|\x_0) ,
\end{equation}
where we used Eq. (\ref{eq:p_int}) to evaluate the integral of
$p(\x,t|\x_0)$ in the first term of Eq. (\ref{eq:U_general}).
Comparing this expression with the integral equation
(\ref{eq:Gs_int}), we reproduce here by changing the integration
variable from $t'$ to $t-t'$:
\begin{align}  \nonumber
G_s(t|\x_0) & = s + \sum\limits_{m=0}^M \int\limits_{\Gamma_m} d\x \\  \label{eq:Gs_intS}
& \times \int\limits_0^t dt' \kappa_m \, p(\x,t-t'|\x_0) \bigl([G_s(t'|\x)]^m - s\bigr).
\end{align}
Subtraction of these expressions yields
\begin{align} \nonumber 
G_s(t|\x_0) & - U(t|\x_0) = \sum\limits_{m=0}^M \int\limits_0^t dt' \int\limits_{\Gamma_m} d\x \, p(\x,t-t'|\x_0) \\
& \times \biggl(\kappa_m [G_s(t'|\x)]^m - f_m(t'|\x)\biggr)  ,
\end{align}
which holds for any $t > 0$ and any $\x_0 \in \Omega$.  If we set
$f_m(t'|\x) = \kappa_m [G_s(t'|\x)]^m$ on $\x \in \Gamma_m$, the
generating function becomes identical with $U(t|\x_0)$, which is the
solution of the initial-value problem (\ref{eq:U_problem}).  We
conclude that the generating function satisfies the backward
Fokker-Planck equation:
\begin{subequations}  \label{eq:Gs_PDEgeneral}
\begin{align}  \label{eq:Gs_FP}
\partial_t G_s(t|\x_0) & = \L^*_{\x_0} G_s(t|\x_0) \quad (\x_0 \in \Omega), \\  \nonumber
\n_{\x_0} \cdot D(\x_0) \nabla G_s(t|\x_0) & = \kappa_m \bigl([G_s(t|\x_0)]^m - G_s(t|\x_0)\bigr)  \\ \label{eq:Gs_FP_BC}
& \hskip 16mm (\x_0\in\Gamma_m), \\  \label{eq:Gs_FP_ini}
G_s(0|\x_0) & = s.
\end{align}
\end{subequations}
This is one of the main results.
For ordinary diffusion with a constant diffusivity $D$, one has $\L =
\L^* = D \Delta$, so that the problem (\ref{eq:Gs_PDEgeneral}) is
reduced to Eqs. (\ref{eq:Gs_diff}, \ref{eq:Gs_BC}).

\section{Distribution of the population size}
\label{sec:Qk}

By its definition, the generating function $G_s(t|\x_0)$ determines
the distribution of the population size $\N(t)$:
\begin{equation}
G_s(t|\x_0) = \E_{\x_0}\{ s^{\N(t)}\} = \sum\limits_{k=0}^\infty s^k \, Q_k(t|\x_0),
\end{equation}
from which the probability $Q_k(t|\x_0)$ of having $k$ particles at
time $t$ reads
\begin{equation}  \label{eq:Qk_def}
Q_k(t|\x_0) =\P_{\x_0}\{ \N(t) = k\} = \frac{1}{k!} \lim\limits_{s\to 0} \partial_s^k G_s(t|\x_0).
\end{equation}
The evaluation of the $k$-th order derivative with respect to $s$ at
$s = 0$ transforms the integral equation (\ref{eq:Gs_intS}) and the
PDE problem (\ref{eq:Gs_PDEgeneral}) to access the probabilities
$Q_k(t|\x_0)$.

In fact, the $k$-fold application of $\partial_s$ to
Eq. (\ref{eq:Gs_intS}) yields after evaluation at $s = 0$:
\begin{align*}
Q_k(t|\x_0) & = \delta_{k,1} + \sum\limits_{m=0}^M \int\limits_{\Gamma_m} d\x \int\limits_0^t dt' \kappa_m \, p(\x,t-t'|\x_0) \\
& \times \biggl[\frac{1}{k!} \bigl(\partial_s^k [G_s(t'|\x)]^m\bigr)_{s=0} - \delta_{k,1}\biggr].
\end{align*}
The sum of the integrals containing $\delta_{k,1}$ can be further
simplified by using the identity (\ref{eq:p_int}).  Likewise, the
$k$-th order derivative of the $m$-th power of $G_s(t'|\x_0)$ is
evaluated by using the generalized Leibniz rule:
\begin{widetext}
\begin{align*}
\frac{1}{k!} \bigl(\partial_s^k [G_s(t'|\x)]^m\bigr)_{s=0} & = 
\frac{1}{k!} \sum\limits_{i_1,\ldots,i_m \geq 0 \atop i_1+\ldots+i_m = k} \frac{k!}{i_1! \cdots i_m!} 
\times \underbrace{(\partial_s^{i_1} G_s(t'|\x_0))_{s=0}}_{=i_1! \, Q_{i_1}(t'|\x_0)} \cdots 
\underbrace{(\partial_s^{i_m} G_s(t'|\x_0))_{s=0}}_{=i_m! \, Q_{i_m}(t'|\x_0)} \\
& = \sum\limits_{i_1,\ldots,i_m \geq 0 \atop i_1+\ldots+i_m = k} Q_{i_1}(t'|\x_0) \cdots Q_{i_m}(t'|\x_0).
\end{align*}
\end{widetext}
We obtain thus the following closed-form integral equation for
$Q_k(t|\x_0)$:
\begin{align} \nonumber
Q_k(t|\x_0) & = \delta_{k,1} S(t|\x_0) + \sum\limits_{m=0}^M \hspace*{-0.5mm} \int\limits_{\Gamma_m} \hspace*{-1mm} d\x 
\hspace*{-1mm} \int\limits_0^t dt' \kappa_m \, p(\x,t-t'|\x_0) \\
& \times \sum\limits_{i_1,\ldots,i_m \geq 0 \atop i_1+\ldots+i_m = k} Q_{i_1}(t'|\x_0) \cdots Q_{i_m}(t'|\x_0),
\end{align}
where
\begin{equation}
S(t|\x_0) = \int\limits_{\Omega} d\x \, p(\x,t|\x_0)
\end{equation}
is the survival probability.  Since this equation expresses
$Q_k(t|\x_0)$ in terms of $Q_0, Q_1,\ldots,Q_k$, one can determine
these probabilities iteratively.  We used this description to compute
numerically the probabilities $Q_k$ shown in Figs. \ref{fig:Qk} and
\ref{fig:Qn_surf2}.

Alternatively, the evaluation of the $k$-th order derivative with
respect to $s$, applied to the PDE problem (\ref{eq:Gs_PDEgeneral}),
yields
\begin{subequations}  \label{eq:Qk_PDEgeneral}
\begin{align} 
& \partial_t Q_k(t|\x_0) = \L^*_{\x_0} Q_k(t|\x_0) \quad (\x_0 \in \Omega), \\     \label{eq:Qk_PDEgeneral_BC}
& \n_{\x_0} \cdot D(\x_0) \nabla Q_k(t|\x_0) + \kappa_m Q_k(t|\x_0) \\ \nonumber
& \quad = \kappa_m  \sum\limits_{i_1,\ldots,i_m \geq 0 \atop i_1+\ldots+i_m = k} Q_{i_1}(t|\x_0) \cdots Q_{i_m}(t|\x_0) \quad (\x_0\in\Gamma_m), \\  
& Q_k(0|\x_0) = \delta_{k,1}.
\end{align}
\end{subequations}
For ordinary diffusion, Eq. (\ref{eq:Qk_PDEgeneral_BC}) is reduced to
Eq. (\ref{eq:Qn_BC}).

\section{Moments of the population size}
\label{sec:Nk}

The generating function $G_s(t|\x_0)$ also determines the $k$-th order
moment of the population size:
\begin{equation}  \label{eq:Nk_def}
N_k(t|\x_0) =\E_{\x_0}\{ [\N(t)]^k\} = \lim\limits_{s\to 1} (s\partial_s)^k G_s(t|\x_0).
\end{equation}
The $k$-fold application of $(s\partial_s)$ to the integral equation
(\ref{eq:Gs_intS}) or to the PDE problem (\ref{eq:Gs_PDEgeneral})
gives access to these moments.
In particular, we obtain the integral equation
\begin{align*}
N_k(t|\x_0) & = S(t|\x_0) + \sum\limits_{m=0}^M \int\limits_{\Gamma_m} d\x \int\limits_0^t dt' \kappa_m \, p(\x,t-t'|\x_0) \\
&  \biggl((s\partial_s)^k [G_s(t'|\x)]^m\biggr)_{s=1} \,,
\end{align*}
where we used again the identity (\ref{eq:p_int}) and the elementary
property $(s\partial_s)^k s = s$.  The $k$-th iterative application of
the operator $s\partial_s$ to the $m$-th power of $G_s(t'|\x_0)$ can
be evaluated by using again the generalized Leibniz rule:
\begin{widetext}
\begin{align*}
\biggl((s\partial_s)^k [G_s(t'|\x)]^m\biggr)_{s=1} & = 
\sum\limits_{i_1,\ldots,i_m \geq 0 \atop i_1+\ldots+i_m = k} \frac{k!}{i_1! \cdots i_m!} 
 \underbrace{(s\partial_s)^{i_1} G_s(t'|\x_0))_{s=1}}_{=N_{i_1}(t'|\x_0)} \cdots 
\underbrace{(s\partial_s)^{i_m} G_s(t'|\x_0))_{s=1}}_{=N_{i_m}(t'|\x_0)} \\
& = \sum\limits_{i_1,\ldots,i_m \geq 0 \atop i_1+\ldots+i_m = k} 
\frac{k!}{i_1! \cdots i_m!} N_{i_1}(t'|\x_0) \cdots N_{i_m}(t'|\x_0).
\end{align*}
\end{widetext}
We obtain consequently the following closed-form integral equation for
$N_k(t|\x_0)$:
\begin{align} \nonumber
N_k(t|\x_0) & = S(t|\x_0) + \sum\limits_{m=0}^M \int\limits_{\Gamma_m} d\x \int\limits_0^t dt' \kappa_m \, p(\x,t-t'|\x_0) \\
& \times \sum\limits_{i_1,\ldots,i_m \geq 0 \atop i_1+\ldots+i_m = k} \frac{k!}{i_1! \cdots i_m!} N_{i_1}(t'|\x_0) \cdots N_{i_m}(t'|\x_0).
\end{align}
Since this equation expresses $N_k(t|\x_0)$ in terms of
$N_1,\ldots,N_k$, one can compute these moments iteratively.  This
recursive structure mirrors the hierarchy of moment equations
encountered in many stochastic processes, while the coupling is
entirely confined to the boundary conditions.

Alternatively, the $k$-fold application of $s\partial_s$ to the PDE
problem (\ref{eq:Gs_PDEgeneral}) yields
\begin{subequations}  \label{eq:Nk_PDEgeneral}
\begin{align}  
& \partial_t N_k(t|\x_0) = \L^*_{\x_0} N_k(t|\x_0) \quad (\x_0 \in \Omega), \\  \label{eq:Nk_FP_BC}
& \n_{\x_0} \cdot D(\x_0) \nabla N_k(t|\x_0) + \kappa_m N_k(t|\x_0) \\ \nonumber
& = \kappa_m  \hspace*{-5mm}  
\sum\limits_{i_1,\ldots,i_m \geq 0 \atop i_1+\ldots+i_m = k}   \hspace*{-3mm}
\frac{k!}{i_1! \cdots i_m!} N_{i_1}(t|\x_0) \cdots N_{i_m}(t|\x_0) \quad (\x_0\in\Gamma_m), \\  
& N_k(0|\x_0) = 1.
\end{align}
\end{subequations}
For $k = 1$, the right-hand side of Eq. (\ref{eq:Nk_FP_BC}) is simply
$m N_1(t|\x_0)$, so that the above nonlinear PDE is reduced to a
linear problem:
\begin{subequations}  \label{eq:N1_PDEgeneral}
\begin{align}  
\partial_t N_1(t|\x_0) & = \L^*_{\x_0} N_1(t|\x_0) \quad (\x_0 \in \Omega), \\   \nonumber
\n_{\x_0} \cdot D(\x_0) \nabla N_1(t|\x_0) & = (m-1)\kappa_m N_1(t|\x_0)  \\ \label{eq:N1_FP_BC}
&  \hskip 16mm (\x_0\in\Gamma_m) , \\  
N_1(0|\x_0) & = 1.
\end{align}
\end{subequations}
As discussed in the main text, Eq. (\ref{eq:N1_FP_BC}) on $\Gamma_0$
is the standard Robin boundary condition for conventional
diffusion-controlled reactions on a partially absorbing region
$\Gamma_0$ with a positive reactivity $\kappa_0$.  In turn, the same
condition with $m \geq 2$ allows one to interpret the effect of
autocatalytic branching events on $\Gamma_m$ on the mean population
size as reaction events with {\it negative} effective reactivity
$(1-m)\kappa_m$.

Since the initial-value problem (\ref{eq:N1_PDEgeneral}) is linear,
standard spectral tools apply directly to analyze its asymptotic
behavior.  In particular, if the spectrum of the backward
Fokker-Planck operator $\L_{\x_0}^*$ is discrete, the principal
(smallest) eigenvalue of $-\L_{\x_0}^*$ determines the long-time
asymptotic behavior of the mean population size: $N_1(t|\x_0) \propto
e^{-\lambda_0 t}$.  The presence of regions with negative reactivity
can lead to a negative value of $\lambda_0$.  As a consequence, the
principal eigenvalue allows one to distinguish three asymptotic
regimes: (i) subcritical ($\lambda_0 > 0$), as for conventional
diffusion-controlled reactions, (ii) critical ($\lambda_0 = 0$) when
the mean population size reaches a steady-state limit, and (iii)
supercritical ($\lambda_0 < 0$) when the mean population size grows
exponentially.  In the main text, we provided the overall qualitative
picture of the population dynamics in these three regimes.

\section{Extension to spatially delimited subpopulations}
\label{sec:extension}

The formalism can be extended to incorporate spatially resolved
population dynamics.  Let $\omega \subset\Omega$ be a subset of the
domain $\Omega$, and we consider the number of particles,
$\N^\omega(t)$, located in $\omega$ at time $t$.  The generating
function $G^\omega_s(t|\x_0) = \E_{\x_0}\{ s^{\N^\omega(t)}\}$ of the
subpopulation size $\N^\omega(t)$ again satisfies the probabilistic
renewal equation (\ref{eq:Gs_prob}) that we rewrite here as
\begin{align} \label{eq:Gs_prob_new}
G^\omega_s(t|\x_0) & = \E_{\x_0}\{ s^{\N^\omega(t)}\, 1_{t = T} \} \\  \nonumber
& + \sum\limits_{m=0}^M \E_{\x_0}\left\{ s^{\N^\omega_1(t-\tau_m) + \cdots + \N^\omega_m(t-\tau_m)} 1_{\tau_m = T}\right\}.
\end{align}
Compared to Eq. (\ref{eq:Gs_prob}), the only difference appears in the
first term: even if the initial particle has not undergone any
reaction up to time $t$ (since $t < \min\{\tau_k\}$), the number of
particles in $\omega$ is either $1$ or $0$.  As a consequence, the
integral equation (\ref{eq:Gs_int}) has to be modified to distinguish
these two options, yielding
\begin{align}  \label{eq:Gs_int_new}
& G^\omega_s(t|\x_0) = s \int\limits_\omega d\x \, p(\x,t|\x_0) + \int\limits_{\Omega\backslash \omega} d\x \, p(\x,t|\x_0) \\ \nonumber
& \qquad + \sum\limits_{m=0}^M \int\limits_{\Gamma_m} d\x   \int\limits_0^t dt' \kappa_m \, p(\x,t'|\x_0) \, [G^\omega_s(t-t'|\x)]^m .
\end{align}
Here, the first term gives the contribution $s^1$ when $\N^\omega(t) =
1$ (with the integral of the propagator over $\omega$ representing the
probability of finding the initial particle in $\omega$ at time $t$),
whereas the second term gives the contribution $s^0$ when
$\N^\omega(t) = 0$.  The other terms remain unchanged.  Using the
identity (\ref{eq:p_int}), we can rewrite the above equation as
\begin{align}  \label{eq:Gs_int_new2}
& G^\omega_s(t|\x_0) = s + (1-s) \int\limits_{\Omega\backslash \omega} d\x \, p(\x,t|\x_0)  \\  \nonumber
& + \sum\limits_{m=0}^M \int\limits_{\Gamma_m} d\x   \int\limits_0^t dt' \kappa_m \, p(\x,t'|\x_0) \, \bigl([G^\omega_s(t-t'|\x)]^m - s\bigr).
\end{align}
This new integral equation differs from Eq. (\ref{eq:Gs_int}) just by
the second term.  In the special case $\omega = \Omega$, this term
disappears, and we recover Eq. (\ref{eq:Gs_int}), as it should be.  In
the same vein, the PDE description (\ref{eq:Gs_PDEgeneral}) remains
valid, except for the initial condition (\ref{eq:Gs_FP_ini}), which is
replaced by
\begin{equation}
G^\omega_s(0|\x_0) = s + (1-s) 1_{\Omega\backslash \omega}(\x_0),
\end{equation}
where $1_{\Omega\backslash \omega}(\x_0)$ is the indicator function of the
complementary set $\Omega\backslash \omega$.

As earlier, the generating function $G^\omega_s(t|\x_0)$ determines
the probability distribution and the moments of the subpopulation size
$\N^\omega(t)$ via Eqs. (\ref{eq:Qk_def}, \ref{eq:Nk_def}).  One can
therefore easily adjust the related integral equations and PDEs to
this setting.  For instance, the PDEs (\ref{eq:Qk_PDEgeneral}) and
(\ref{eq:Nk_PDEgeneral}) for $Q_k(t|\x_0)$ and $N_k(t|\x_0)$ still
hold, except for the initial conditions, which now read as
$Q_k(0|\x_0) = \delta_{k,0} 1_{\Omega\backslash \omega}(\x_0) +
\delta_{k,1} 1_{\omega}(\x_0)$ and $N_k(0|\x_0) = \delta_{k,1}
1_\omega(\x_0)$, respectively.

The same procedure can be applied to other observables, illustrating
the flexibility of the present framework.

\end{document}